# Wrinkling and crumpling in twisted few and multilayer CVD graphene: High density of edge modes influencing Raman spectra.

D. Nikolaievskyi[1,3], M. Torregrosa[1], A. Merlen[2], S. Clair[2], O. Chuzel[3], J.-L. Parrain[3], T. Neisus[4],

A. Campos[4], M. Cabie[4], C. Martin[1], C. Pardanaud[*1]

[1]*Aix Marseille Univ, CNRS, PIIM, AMUTech, Marseille, France*
[2]*Aix Marseille Univ, Université de Toulon, CNRS, IM2NP, Marseille, France*
[3]*Aix Marseille Univ, CNRS, Centrale Marseille, iSm2, AMUTech, Marseille, France*
[4]*Aix Marseille Univ, CNRS, Centrale Marseille, FSCM, Marseille, France*

**Graphical abstract**

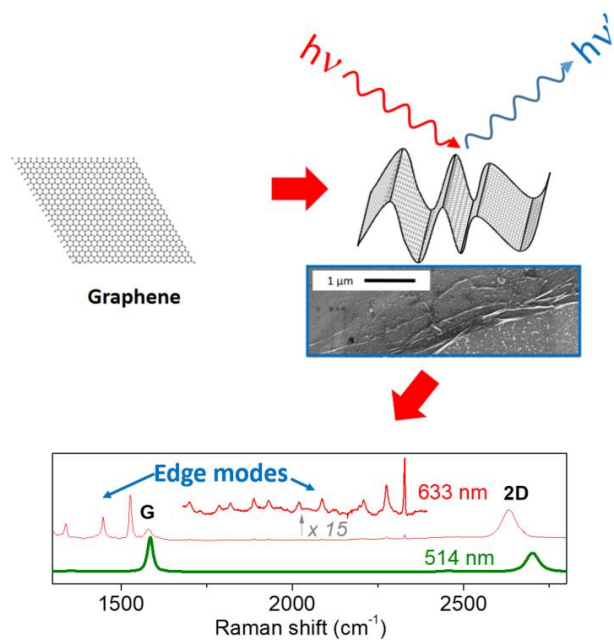

*Corresponding author. Tel: +334 13 94 64 59. E-mail: cedric.pardanaud@univ-amu.fr

**Abstract**

Richness and complexity of Raman spectra related to graphene materials is established from years to decades, with, among others: the well-known G, D, 2D,... bands plus a plethora of weaker bands related to disorder behavior, doping, stress, crystal orientation or stacking information. Herein, we report on how to detect crumpling effects in Raman spectra, using a large variety of few and multilayer graphene. The main finding is that these crumples enhance the G band intensity like it does with twisted bi layer graphene. We updated the D over G band intensity ratio versus G band width plot, which is generally used to disentangle point and linear defects origin, by reporting surface defects created by crumples. Moreover, we report for the first time on the existence 23 resonant additional bands at 633 nm. We attribute them to edge modes formed by high density of crumples. We use Raman plots (2D bands versus G band positions and widths) to gain qualitative information about the way layers are stacked.

## 1. Introduction

Graphene, due to its 2D nature, possesses charge carriers that behave as massless Dirac fermions, with a linear dispersion energy curve with a Fermi velocity close to $10^6$ m/s [1]. Most of the astounding properties and potentialities heard about single layer graphene (SLG), starting with its semi metallic nature, come from zones close to the K and K' points of the momentum space, the so-called Dirac cones [2-4]. However, graphene is not perfectly flat, but corrugated. Depending on its preparation method, graphene can also be crumpled or contain wrinkles [5-7]. This can cause strain, curvature [8] or enhance chemical reactivity [9]. However, this imperfect graphene became a subject of numerous studies leading to the emergence of exciting properties: modifying graphene by introducing strain or curvature will influence electronic properties. For example, it has been found that strain can induce a pseudo magnetic field that acts on charge carriers as a real giant magnetic field does, as proved by the presence of quantized Landau levels (LL) [10]. They can also be induced by the presence of wrinkles [11-13]. This is an ongoing building field of material science that contributes to bridge the gap between condensed matter physics and quantum field theories [14]. Another way to modify graphene and tune its electronic properties [15] is to stack several layers [16, 17], playing with rotational disorder [18]. Twisted bi layer graphene (tBLG), few layer graphene (FLG) and multi layer graphene (MLG) with a twist became model systems for strongly correlated electrons [19], exhibiting superconductivity [20],[21] and other behaviors close to the 1.1° so called *magic angle* [22]. A key point in that phenomena understanding is the creation of a periodic modulation of the potential by means of superlattices [23] [24] obtained by twisting two layers [25], layers with a twisted interface between two bilayers [26], or more recently for multiple interfaces in MLG [27]. Interaction between Dirac cones from different layers generate van Hove singularities (vHs) [28], enhancing the carrier density of state, with corresponding energies falling interestingly in the visible spectral region. It opens new opportunities in optoelectronics [29], as these gaps can be tuned [30]. vHs were primary thought to appear only for small angles (<5°) where the layers are in strong interaction, and not for large angles (20-30°) where the layers are nearly decoupled from each other. Later, vHS were also observed for large angles and even for MLG [31]. Environmental effects [32], strain [33] as well as electrostatic gating [29] can modify vHs behavior as well, offering extra opportunities of tuning. Additionally, the introduction of defects in these superlattices like local defects (vacancies and Stone-Thrower-Wales defects [34]) or macroscopic defects (wrinkles [35]) is also of interest. Moreover it was shown that bilayer graphene deposited on a corrugated surface displays both vHS and pseudo magnetic field, coexisting without mutual disturbance [36]. This last result was obtained by interpreting Raman microscopy data. The purpose of this paper is to focus on the complexity of Raman spectra of crumpled/wrinkled few and multi layer graphene.

Raman microscopy, an optical technique, is one of the most powerful tools to investigate graphene-based materials and their modifications. Far from the research field, one would quickly conclude that for modified graphene, there are only three bands in the spectra, and interpretation is easy. Taking a closer look reveals that the story is more complex. First, it is an inelastic scattering process involving mainly phonons, but also charge carriers [37, 38]. Most of the time, only phonons at the center of the Brillouin zone are probed using Raman spectroscopy, which is the case for longitudinal and transverse optical phonons, with a band called the G band (1582 cm$^{-1}$ for SLG). Multiple resonance mechanisms give rise to the so called D band (1350 cm$^{-1}$ with 514 nm laser) and 2D bands (2680 cm$^{-1}$ with 514 nm laser) [39, 40], involving phonons and/or defects together with charge carriers close to the Dirac cones. These resonances occur as intravalley or intervalley (inner and outer) processes. G, D and 2D bands is the trio of bands introduced few lines above. Thus, Raman spectrum can be very rich due to a bestiary of bands related to different phenomena. First, the D band is sensitive to defects [41], whereas the G and 2D bands are sensitive to doping and strain together [42], interaction between graphene layers (2D band shape is found featured when MLG is well staked, and Lorentzian-like when it is not [43-45] [46], with variability in its width depending on the number of layers, but not limited to this parameter). Double resonance mechanism was also found to give rise to weak signatures below 900 cm$^{-1}$ [47] [48] in the range 1300-1700 cm$^{-1}$ [49] and in the 1700-2000 cm$^{-1}$ spectral region, due to turbostraticity [50, 51]. In tBLG, other bands activated by either intralayer or interlayer processes [52], and involving different phonon branches could be observed, allowing the Brillouin zone to be partially explored by tuning the twist angle [53-57]. Influence of defects in tBLG has been investigated as well [58]. Like the tBLG system, two domains composed of N layers and M layers with a twist angle at their interface is also of interest and Raman spectroscopy is found useful~~l~~ [59]. It is generally referred as N+M configuration. Electronic resonance scattering of twisted 1+1, 1+2 and different 1+3 layers configurations could also lead to very sensitive Raman bands that could be used in the future to increase the knowledge of how MLG layers are stacked, as previously calculated [60]. The spectroscopic parameters of these bands could be affected by external constrains like the formation of wrinkles which could affect band intensities thanks to the formation of vHS [61] [62]. As we will see in the next section, we complete ~~this data of bands~~ knowledge, showing that crumpling / wrinkling could modify band intensities and ~~also~~ enhance edge vibrational modes in the case of MLG.

To guide the reader, we sum up here the main findings. We have analyzed the Raman data of several FLG or MLG obtained from commercial graphene transferred onto a modified silicon substrate. We focused on crumpling behavior and stacking behavior. Our main findings and observation, demonstrated later in the text, are resumed here:

(i) Crumpling or wrinkling enhances the G band intensity by up to two orders of magnitude, which modifies the D over G band ratio, leading to a wrong appreciation of defect behavior

(ii) This enhancement has common origin with the G band enhancement found for twisted bi layer graphene

(iii) We could add a new category of defects in the Cançado plot reporting D over G band intensity ratio in function of G band width: "2D defects"

- (iv) These "2D defects" are caused by the two surfaces of a crumple that are facing each other with a random twist angle
- (v) Simultaneously, we report the observation of 23 bands which are more intense when samples contain more crumples
- (vi) These bands are only observed at 633 nm, with a resonance mechanism to identify
- (vii) 12 of these 23 bands are observed in the range 600-1600 cm$^{-1}$. The 11 other bands are observed at higher wavenumber. Their wavenumber could be found as second harmonic or a combination of the 12 wavenumbers cited previously.
- (viii) Lee plot (displaying 2D band position in function of the G band position) are reinvestigated for crumpled FLG and MLG.
- (ix) By creating MLG from FLG with different initial doping amount, we evidence a memory effect, like it was observed for twisted bi layer graphene
- (x) 2D band shape of our MLG and FLG samples is single Lorentzian
- (xi) Plotting 2D band width in function of G band width reveasl our samples are intermediated between turbostratic graphite and twisted bi layer graphene

In section 2, we describe the experimental methods. In section 3 we describe the 6 kinds of FLG and MLG samples we used. In section 4 we give results by first discussing general Raman trends, then focusing on the 23 new bands and finally by analyzing the influence of crumples on the classical Raman spectra.

## 2. Experimental methods

*Raman microscopy*

Raman spectra were obtained using a Horiba Jobin Yvon HR800 setup with excitation wavelengths of $\lambda_L$ = 514 and 633 nm, a ×100 objective (numerical aperture of 0.9), a 600 grooves/mm grating. 457, 488 and 785 nm were used only on one of the samples for testing better the resonance behavior of some new bands reported in this paper (see supplementary information). The resolution was about 1 cm$^{-1}$. Band positions were systematically checked using atmospheric $N_2$ and $O_2$ vibrations. Power was kept below 1mW.µm$^{-2}$ to prevent from damages. Tests measuring first at 0.1mW.µm$^{-2}$, followed by a measurement with 1mW.µm$^{-2}$ and a final one at 0.1mW.µm$^{-2}$ were performed to ensure no irreversible heating/degradation effects. Polarization vector was set in the wafer plane. As graphene layers are crumpled in random directions at the lateral micron scale, it means that we could not retrieve easily an information related to polarization, as shown in supplementary information. Then, we did not select systematically a polarization direction just before the detector, except on one dedicated experiment to control the polarization behavior of some new bands reported in this paper.

*FLG and MLG preparation*

Graphenea® "Easy Transfer", a CVD (chemical vapor deposition) graphene protected by a sacrificial poly(methyl methacrylate) (PMMA) layer has been used to form FLG and MLG presented in this study. By a simple introduction into distilled water, the PMMA/graphene film is released from the polymeric substrate which is immediately taken out and thrown off. The PMMA/graphene film remains floating on the water surface and is then "caught" and transferred upon a dedicated surface. Then the sample is kept intact for one hour, before being heated at 150°C under inert atmosphere for better adhesion of graphene to the surface. Finally, the PMMA layer is removed by acetone and isopropanol (IPA) treatments. Dedicated surface are silicon wafer with native oxide. Flat surface leads to pt-FLG, wr-FLG and cr-FLG, which are respectively pristine, wrinkled or crumpled few layer graphene layers, as defined in table 1. Not flat surface (laser machining superimposed micrometric disks of the surface, see Figure S1) leads to od-FLG and nd-FLG, od standing for "on disk" and nd for "near disk". To create these superimposed disks, silicon wafers were machined under ambient conditions in air using a 355nm pulsed laser (Laser Micromachining Ltd., UK). HRTEM measurements made on a thin foil cut on that disks haves shown no increase of the surface oxide thickness. Topography shows small as well as strong height gradients as can be seen in SI. fd-MLG, fd standing for "folded" , have been obtained by using our cut and fold process on pt-FLG. The method, consisting in zig zag motion of the AFM tip to draw squares at a constant vertical force, was described earlier [63]. There were 512 lines on 4 µm, which represents the length of the squares at a speed of 8 µm s$^{-1}$.

*Electron microscopies*

Scanning electron microscopy (SEM) was performed with a Zeiss Gemini 500 electron microscope equipped with a secondary electron detector, ideal for displaying surface structures. Secondary electron images were acquired at low accelerating voltages (1-5 kV range) to enhance topographic contrast. To obtain a thin cross-sectional view of the graphene multi layer a thin lamella was prepared by focused ion beam milling (FIB) using a FEI Heilos 600 Nanolab. The transmission electron microscopy (TEM) characterization of the FIB lamella was carried out using a FEI TITAN 80-300 electron microscope equipped with an image aberration corrector. The acceleration voltage of the instrument was set to 200kV.

## 3. Sample description

Samples were obtained by transferring commercial CVD graphene with standard wet methods onto a Si wafer on which alignments of disk-like structures (diameter 10 μm and height 600 nm) were produced by laser machining. The transferred graphene was studied using SEM, TEM and Raman and was identified as FLG [56]. Fig. 1 shows images of the six types of zones, defined at the scale of the Raman laser spot (micrometric), that were probed. These are:

(i) zones called pt-FLG where SEM images show a limited amount of wrinkling or crumpling (Fig. 1a and 1j); TEM image of Fig. 1b shows that pt-FLG contains 2 or 3 layers. Observations made on 50 TEM images and probing a length of about 15 μm show most often 2 or 3 layers (rarely up to 7 in some places).

(ii) zones called wr-FLG where SEM images show significant wrinkling (Fig. 1a) ;

(iii) zones called cr-FLG where SEM images show a high degree of crumpling (Fig. 1l), consecutive to a failure in the transfer process;

(iv) zones called od-FLG situated on top of the laser machined disks. The disks are asymmetric and curved (see the profiles obtained by confocal microscopy in SI). On top of the disks, graphene was ripped in the center and localized mainly on the edges (Fig. 1h); wrinkles were visible (Fig. 1k). The corresponding G band was found very intense, up to two orders of magnitude higher than for pt-FLG (Fig. 1i).

(v) zones at a few micrometers in the neighboring of the disks (called nd-FLG, expected to be intermediate between pt-FLG and od-FLG (Fig. 1h). Actually, nd-FLG was less ripped than od-FLG, with elongated holes perpendicular to the disk alignment. The corresponding G band intensity (Fig. 1i) was intermediate between those of od-FLG pt-MLG (Fig. 1i), showing that the topography created by the disk alignment plays a role as far as 5-10 μm away.

(vi) Finally, MLG was created using an AFM cut and fold method described in [56]: scanning with the AFM tip in contact mode square zones (typically 4x4 μm$^2$) of pt-FLG with forces ranging from 4 to 8 μN locally scratches the graphene layer while folding it along the slow-scan direction [63]. Fig. 1c and 1e show SEM images of scratched and folded zones. Fig. 1d is the Raman map corresponding to Fig. 1c, the zero G band intensity (in blue) showing that graphene was scratched within the scanned zone and the maximal G band intensity (in red) showing that graphene was reassembled at the top of the scanning zone on a typically 4 μm x 100 nm zone (called fd-MLG). Fig 1f and 1g are TEM images showing the folding process as leading to well-ordered stacking combined with a messy 3D organization. The number of stacked layers was found to be 3 to 20, with an interlayer distance of typically 0.36 nm, like with FLG.

| Sample | Description |
| --- | --- |
| pt-FLG | FLG probed on a zone with limited wrinkling or crumbling, named pristine |
| wr-FLG | FLG probed on a zone with significant wrinkling |
| cr-FLG | FLG showing significant crumpling due to transfer failure |
| od-FLG | FLG probed on the laser machined disks of the wafer (od = on-disk) |
| nd-FLG | FLG probed in the vicinity – 10 μm – of the disks (nd = near disk) |
| fd-MLG | MLG obtained from pt-FLG using the AFM cut and fold method (fd= folded) |

*Table 1.* *Description of the different samples and their production method. They are all obtained from commercially available graphene transferred with the standard wet method onto a Si wafer.*

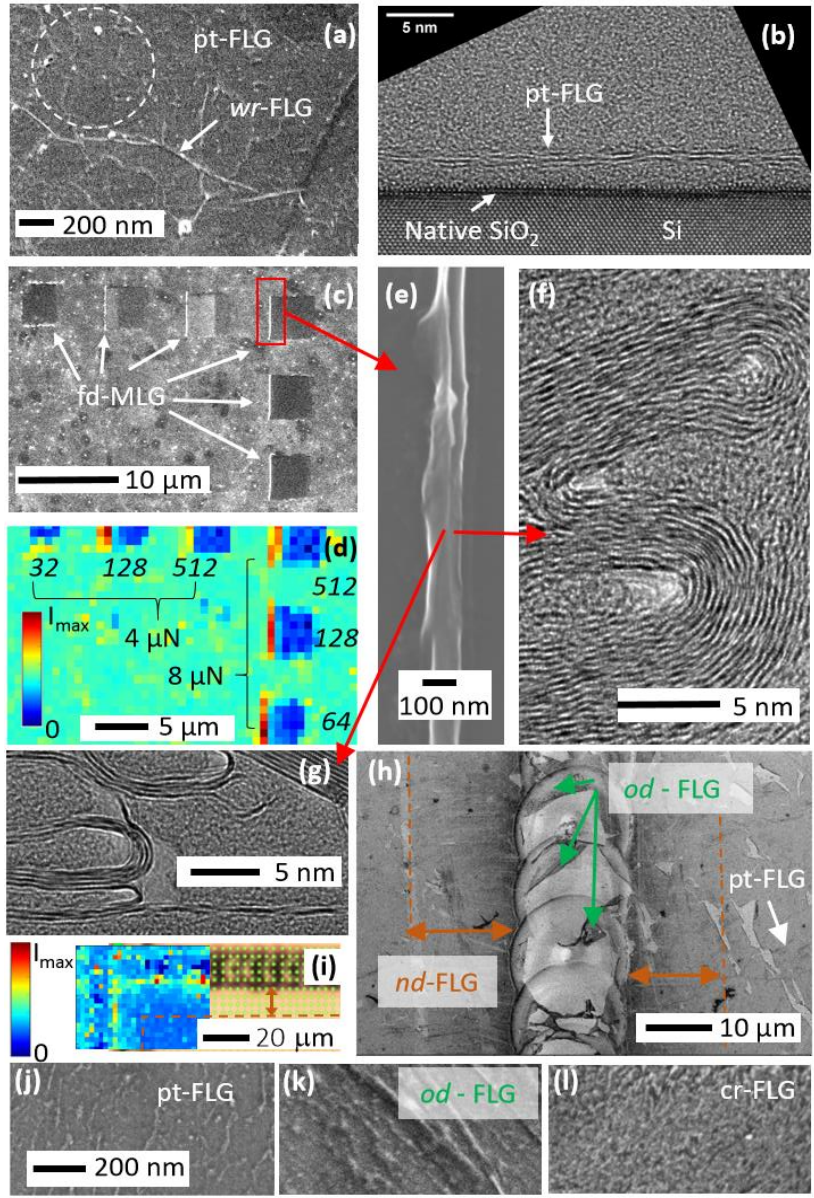

*Figure 1.* FLG and MLG sample presentation. (a) SEM top view and (b) TEM cross section image of pristine FLG, showing a wrinkled zones in (a) ; (c, e) SEM top view images and (d) Raman G band intensity map of fd-MLG ; (f, g) TEM cross section images of fd-MLG ; (h) SEM top view image of FLG on and near disks ; (i) Raman G band intensity map on and near the disks. In (d) and (i) dark blue means no graphene, red corresponds to the most intense G band [63]. (j, k, l) SEM top view images of pt-FLG, od-FLG and cr-FLG, respectively (same scale).

## 4. Results and discussions

*4.1 General Raman characteristics of FLG and MLG*

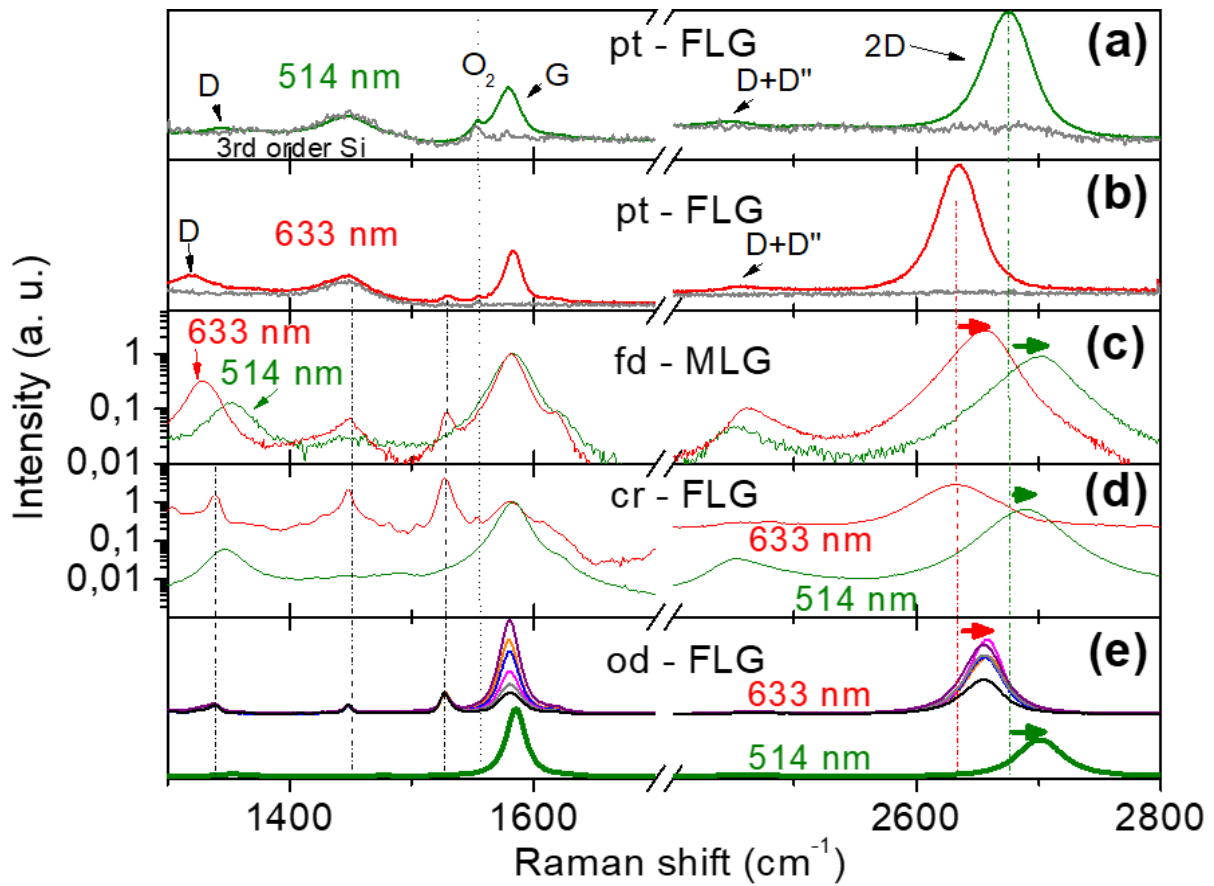

***Figure 2.*** *Multi resonant effects in the Raman spectra of our FLG and MLG samples. Two laser wavelengths are used 514 nm and 633 nm. (a) and (b) are Raman spectra of the same pristine FLG with 514 and 633 nm, respectively. (c) Folded MLG using 514 and 633 nm. (d) Crumpled FLG using 514 and 633 nm. (e) On disk FLG spectra using 514 and 633 nm. Intensity is given with a log scale for subgraph c and d to better highlight some bands.*

Figure 2 displays typical Raman spectra of the samples under study, using two laser wavelength (514 and 633 nm), to investigate multi resonance effects with electronic structure. Spectra with 3 other wavelengths are given in supplementary material for the cr-FLG sample. The term "multi" refers here to the fact that several behaviors in the electronic structure, involving different kinds of electrons, could lead to resonances. For example, electrons involved in van Hove singularities mentioned in the introduction and electrons involved in the double resonance process leading to the D and 2D bands will not be the same electrons and will lead to different effects in the Raman spectra. Fig.2-a and b display spectra of our pristine FLG for respectively 514 and 633 nm lasers and compared to silicon substrate spectrum. Atmospheric $O_2$ and $N_2$ (not shown) are clearly visible at 1530 and 2330 cm$^{-1}$. Third order silicon is found as broad overlapped bands, with the most intense band at 1450 cm$^{-1}$. D band, more intense with the 633 nm than the 514 nm, as anticipated [64], is found overlapped with third order silicon. G, D+D" and 2D bands are clearly seen. The 2D band is downshifted by ≈100 cm$^{-1}$/eV, and D band by half, as expected due to double resonance mechanism when changing the laser wavelength. *fd*-MLG spectra recorded at 514 and 633 nm are displayed in fig.2-c. There is roughly a one order intensity enhancement compared to the pristine FLG, partially due to the stacking under the laser beam, as discussed previously [63]. A very low intense band is found at 1529 cm$^{-1}$ with the 633 nm laser on pt-FLG (nothing is seen at this wavenumber with 514 nm). If a subtraction is made with the silicon related spectrum, the difference spectrum also displays a very low intensity band at 1448 cm$^{-1}$, very close to the 1450 cm$^{-1}$ band from third order silicon. These two bands are found more intense in fd-MLG recorded with 633 nm whereas they are still silent with 514 nm. These two bands are found much more intense in od-FLG, with 633 nm. 514 nm laser sometimes displays a very low band close to 1470 cm$^{-1}$. The od-FLG samples is accompanied by 23 other bands detailed in figure 4. In figure 2-e, we display some spectra recorded at different location of the sample and normalized to the 1448 cm$^{-1}$ band height. Performing this, we could see that the 1302, 1339, 1448 and 1529 cm$^{-1}$ bands (and the weaker ones outside the spectral range of this figure) always superimpose. This is not the case with the G, D and 2D bands which could vary relatively from one location of the sample to the other. From this observation, we could conclude that the set of new bands evolve as a block, and differently than the traditional D, G and 2D bands. First, both G and 2D bands behavior was analyzed. *fd*-MLG and od – FLG 2D bands are blue shifted by ≈ 25 cm$^{-1}$ compared to pristine FLG sample for both 514 and 633 nm wavelengths. A blue shift has been observed when increasing the number of layers [65]. This could reveal a Fermi velocity reduction, or an outer enhanced Raman scattering process involving Dirac cones [66], both based on a double resonance mechanism [67].

*4.2 Information about doping, strain and stacking thanks to the 2D and G band behaviors*

In figure 3-a, 2D band position ($\omega_{2D}$) in all samples is plotted in function of the G band position ($\omega_G$). This plot is generally made for single layer graphene, and called a Lee diagram [42]. It allows estimating simultaneously strain and doping [42, 68]: a slope of 0.75 refers to charge doping variation whereas a slope of 2.2 refers to strain variation, and a combination of both leads to occupy the purple or pink areas. This plot is used to compare qualitatively FLG and MLG samples, taking advantage to the fact that the strain/doping information is stored in FLG/MLG, according to [65]. FLG/MLG samples are compared to the disordered carbons reference that is thought to mimic turbostratic graphene[2]. Three pristine FLG are used, one refers as undoped and two refer as doped (of the order of ≈1-2.10$^{12}$ cm$^{-2}$, one with negative strain, ε, and one with a positive strain, according to Raman analysis). Folded MLG obtained after the cut and fold process of these three pristine MLG are situated in the green region of fig.3-a. Data are spread mainly vertically for the undoped starting sample (tiny G band down shift and huge 2D band blue shift, as reported in [63]), mainly vertically for the initially doped sample with $\varepsilon > 0$ and obliquely for the initially doped sample with $\varepsilon < 0$. The three kinds of fd-MLG do not overlap in the $\omega_{2D}$ vs $\omega_G$ plots, showing that folded layers contain memory of the strain and doping which were present in the initial pt-FLG used to obtain fd-MLG. They spread a large part of the green zone. wr-FLG are found close to the maximal value of the doped fd-MLGs. wr-FLG are locally folded graphene layers and then it is reasonable to note that wr-FLG and fd-MLG have a common behavior. On the same plot notice that end-FLG are spread close to a line parallel to the doping line but upshifted by a systematic offset of ≈10 cm$^{-1}$. Qualitative interpretation of this sample is that doping is heterogeneous (one order of magnitude in amplitude) whereas strain is constant. For od-FLG, several measurements are found grouped in the green zone of fig. 3-a, with high $\omega_{2D}$ value. Then, most FLG/MLG samples, except the pristine FLG, are spread in the green zone. Additionally, the upper edge of the green zone draws a line which is parallel to the doping line, with a systematic 22 cm$^{-1}$ upshift for $\omega_{2D}$. For comparison, the data are related to twisted BLG with various twist angles found in the literature [71, 72]. A blue shift in the 2D band is observed, which is related to the twist angle value, as data are spread in the green region of figure 3-a. There is also a systematic 4 cm$^{-1}$ shift for $\omega_G$ comparing data from Kim [64] and Gadelha [65]. This shift is comparable to the one we obtained for fd-MLG built from undoped and doped pt-FLG. The data obtained by Kim *et al.* show that their samples are probably doped. This is in agreement with band shifts reported when doping is in the 10$^{12}$ charge per cm$^2$ range [73]. This doping could be explained because of the use of TEM grids, as this is the case for our suspended SLG we display in figure 3-a and which is highly doped compared to pt-FLG, supported on Si. To conclude on this paragraph, obtaining at the same time quantitative strain and doping values plus information about stacking using this $\omega_{2D}$ vs $\omega_G$ plot is very tricky for FLG/MLG samples, even if we could observe strong qualitative effects. Interpreting this plot will need more fundamental studies like the one recently published in [74, 75], focused on BLG. Note that the role of vHS, important for BLG as presented in the introduction, has also been pointed out for a higher number of layers. In this study focused on highly oriented pyrolytic graphite (HOPG), it has been shown that the top layer can govern the Raman signature, due to resonance mechanism involving vHS created by the top layer interaction with the staked layers situated below [76]. Then, interpretation on this plot remains complex. One issue to allow starting quantitative interpretation is to prove Fermi velocity is systematically related to a 2D band blue shift[3]. If this is proven, 2 to 23% reduction deduced from Raman spectroscopy [67], [77] will be comforted and could be extended systematically.

---

[2] There is a lack of data in literature about reporting simultaneously $\omega_G$ and $\omega_{2D}$ for turbostratic graphite. We obtained our samples by heating (from 1500°C to 2500°C) a large variety of defective samples [69] [70] . As discussed in figure 6, where Raman data are available simultaneously so that we can report them on graphs, they are taken as behaving similarly to turbostratic graphite.

Figure 3-b displays a bandwidth based plot ($\Gamma_{2D}$ vs $\Gamma_G$). Here we assume a single Lorentzian profile for fitting the bands. As discussed previously [63], most of the FLG/MLG

---

[3] This is not yet the case, there is still a debate, with two possible interpretations. See comment above in the text.

layers are spread around a linear relation of slope 2, that we call 2d-line. Note that this relation has been found previously in a paper where varying the local strain was done, samples with local strain being found at $\Gamma_G \approx 17$ and $\Gamma_{2D} \approx 40$ cm$^{-1}$, whereas it was found at $\Gamma_G \approx 5$-10 and $\Gamma_{2D} \approx 20$ cm$^{-1}$ for unstrained samples [78]. A log-log scale plot displays one decade variation on both axes. wr-FLG, cr-FLG and od-FLG are found close, nd-FLG are found slightly higher in both widths. The memory effect highlighted previously in the $\omega_{2D}$ Vs $\omega_G$ plot for *fd*-MLG is not visible here: in the $\Gamma_{2D}$ Vs $\Gamma_G$ plot, all the *fd*-MLG data are overlapped. This overlap means the information to gain here is of different nature than related to strain and doping and has something to do with interaction between graphene layers, and electron-phonon coupling. For comparison we also display other samples: the set of samples labelled disordered C and discussed at figure 3-a well reproduces widths of turbostratic samples found in the literature [79]. Twisted BLG data come from two papers [71, 72], as for figure 3-a, for comparison. We can now comment the cloud of points: close to $\Gamma_G$ =25 cm$^{-1}$ and $\Gamma_{2D}$ =50 cm$^{-1}$, there is a peculiar branching point connecting the 2d-line with a 3d-line, representative of samples that display 3d stacking order (ABA for example). All the FLG/MLG produced in our study are situated around the line of slope 2, below or close to that branching point. It could mean that layers are "electronically decoupled". This could be counterintuitive if we look again at figure1-f where large domains of up to 20 layers are found well stacked for fd-MLG. But this apparent discrepancy has been pointed out in [75] which focused the study on AB-BLG and twisted BLG: we could regard twisted FLG as electronically decoupled individual stacked layers. What about differences we observed in figure 3-a about twisted BLG? For twisted BLG with an angle increasing up to the magic angle at 1.1°, from [72], both $\Gamma_G$ and $\Gamma_{2D}$ increase nearly linearly with twist angle increasing up to 1.1°. For higher angles, $\Gamma_{2D}$ does not evolve and $\Gamma_G$ diminishes. For angles smaller than 1.1°C, data are outside of the power law drawn by all other samples (referred as "small twist" in the figure), whereas data corresponding to 1.1° falls very close to the values of $\Gamma_G$ and $\Gamma_{2D}$ obtained for some *fd*-MLG and *wr*-FLG. Data concerning tBLG obtained by Kim *et al*, which is potentially doped due to the influence of a gold TEM grid, as discussed earlier, display roughly the same behavior as the one of Gadelha, except data are more spread. Again, this spreading could be explained partially by the electrical doping that broadens and enhances the 2D+ component coming from the double resonance outer process at the K point [80]. Finally, we can resume that there is a power law linking 2D and G bandwidths. Its nature must be understood but the relation is robust. Doping and strain do not affect too much this law[4]. However, this law is found more sensitive on the way layers are stacked.

---

[4] It is assumed that polymer assisted wet transfer used in this study has no influence on the effects reported previously. As a proof, some measurements were displayed from [81]. and graphene was used with and without polymer. It appears that the use of PMMA does not keep away the data from the power law. This power law is then intrinsic, and its origin needs to be understood. As a reference, measurements of a suspended monolayer graphene on 2 µm gold TEM grid, were also performed, corresponding data interpreted as slightly p-doped [82].

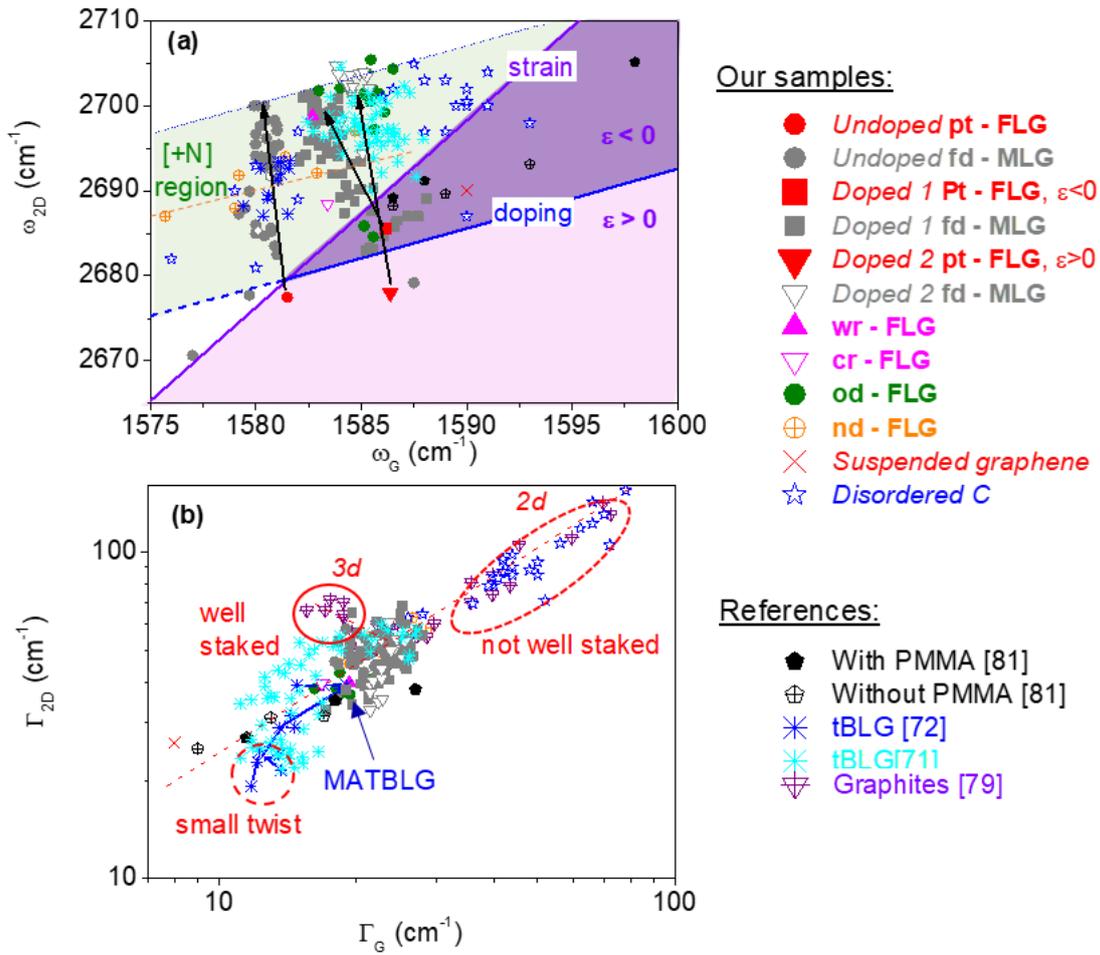

***Figure 3.*** *(a) Doping and strain characterization with a $\omega_{2D}$ vs $\omega_G$ plot. Straight lines correspond to the behavior under doping (blue) or strain (violet). Purple and pink zones correspond to respectively negative and positive strains. Green zone is in practice not accessible for single layer graphene, but is accessible if the number of layer increases (labelled [+N] region here). (b) Structural characterization*
*with a $\Gamma_{2D}$ vs $\Gamma_G$ plot. Wu et al data, corresponding to wet transfer steps influencing graphene are from [81]. Samples labelled Disordered C behave as turbostratic graphite. tBLG in cyan color come from [71] and blue color from [72]. Note that a 45 cm-1 shift was added to $\omega_{2D}$ because of 2D band dispersion for tBLG data from [72] as they were recorded at 633 nm and our plot is relevant for 514 nm.*

## 4.3 Edge state evidence and consequences on the overall Raman spectrum

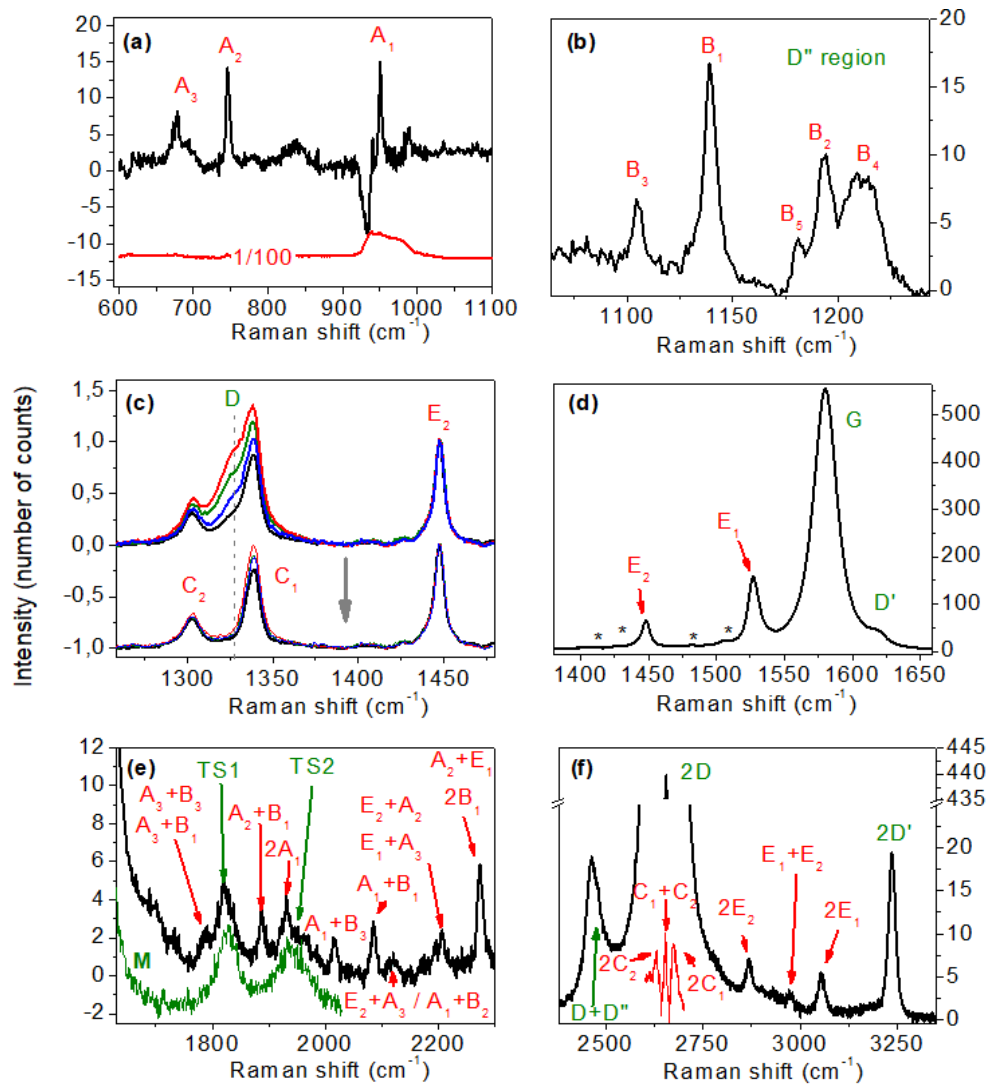

*Figure 4.* Raman spectra recorded at 633 nm for od - FLG. (a) G band, (b) low wave number, (c) D" band region and (d) D band spectral regions, (e) 2D band region, (f) 1700-2100 cm$^{-1}$ TS region. The green spectrum is a reference spectrum. Labelling in green refers to bands due to double resonance mechanism. Depending on the spectral window, we label new bands with A, B, C and E (not D, which is used for some bands related to double resonance mechanism), sorting with an index that increase with a band height diminution. For subfigure (b), this is a difference spectrum with silicon as in this region Si displays a strong and broad second order signature (spectrum divided by 100, without this difference is displayed in red). Upper spectra of subfigure (d) represents raw data. Lower spectra of subfigure (d) represents data with a D band subtracted in between $E_1$ and $E_2$.

Figure 4 displays a typical Raman spectrum of *od* - FLG recorded at 633 nm. Superposed to typical bands supposed to be due to usual double resonance mechanism (G and D' in fig.4-a, D+D", 2D and 2D' in fig.4-e, turbostratic bands TS1 and TS2, also referred as LO+TA and LO+LA respectively *[50, 83, 84]* ), 23 other bands are present. These bands are not seen with 514 nm nor 785 nm, meaning they result from a resonant process, potentially due to van Hove singularities, as presented in the introduction. We label each of that bands observed at 633 nm by a letter plus a number sorted in height (1 is the most intense in the spectral range considered). A is for 600-1000 $cm^{-1}$, B is for 1100-1250 $cm^{-1}$, C is for 1250-1450 $cm^{-1}$, E is for 1400-1650 $cm^{-1}$ ranges, respectively, (letter D is skipped because it is used for the double resonance mechanism induced bands). In fig. 4-e and 4-f, we labelled all the bands by naïvely combining the frequencies of the bands found in the range 600-1650 $cm^{-1}$. Doing that, no anharmonic effects are assumed. Astonishingly, most of the bands are only downshifted by few wavenumbers compared to the expected band position with a harmonic approximation. Only few bands display a larger 30 $cm^{-1}$ downshift compared to the harmonic approximation. Then, we could argue that all the new bands higher than 1650 $cm^{-1}$ could be due to combination modes and overtones. Moreover, the bands arising from combining $C_1$ and $C_2$, should fall in the range of the very intense 2D band, with poor chance to observe something. Nevertheless, by fitting the 2D band with two Lorentzian (to consider inner and outer contributions in Raman signal), and looking on the residue, displayed in fig.4-e, the three corresponding combination bands appear, with widths like the one of the other combination bands. It means that this residue could be attributed to spectroscopic signatures and not by an inappropriate choice of the fitting model. Weaker bands rise close to the $E_1$ and $E_2$ bands, as seen in figure 4-d and S6. Below, we examine all the possible interpretations about the 23 bands source. In brief, among all the physico-chemical hypothesis, some of that bands could potentially be attributed to *i)* PMMA traces (used during the wet transfer method), *ii)* edge phonon modes of graphene, *iii)* one (or more) twisted interface between two graphene layers involving superlattices, *iv)* stacking disorder in multi layer graphene, *v)* electronic resonance signature, *vi)* presence of typical defects such as Stone Wales (SW) and/or divacancies (DV). First, PMMA residue should give rise to signatures mainly at 596, 813, 1453, 1729 and overlapped bands in the range 2951-3002 $cm^{-1}$ (see the corresponding spectrum in SI), the more intense being at 813 and 2951 $cm^{-1}$, whatever the wavelength of the laser used in this study. Only the band at 1453 $cm^{-1}$, and in a lower extent the low intensity band at 2979 $cm^{-1}$, could fit with one of the 23 bands we observed. We have never observed other PMMA related bands, so we could exclude PMMA as a possible origin for these bands. Second, edge phonons in Graphene Nano Ribbons (GNR) are reported many times to give rise to two bands at 1450 $cm^{-1}$ (zigzag edge) and 1530 $cm^{-1}$ (armchair edge) and in a lower extent to bands close to 1130 and 1210 $cm^{-1}$ [85-87]. Bands closely related to these frequencies, mainly at 1449 and 1528 $cm^{-1}$, and also at 1140 and at ≈1200 $cm^{-1}$ are observed. Moreover, theoretical calculations related to edges [88] suggest the existence of low lying vibrational modes that could be compatible with the A bands detected. Even if not all the bands observed have been previously predicted or reported, the global good match with edge vibrational signatures make this hypothesis a good candidate. Third, 677, 746, 949, 1449, 1528 cm$^{-1}$ signatures, could also be interpreted to a ≈20° twist angle between two layers. This involves superlattices selecting phonons far from the Γ point, and could be attributed to TA, ZO, LA, TO and LO phonon branches, respectively [54]. Bands close to 1140 and at ≈1200 cm$^{-1}$ or higher than 1800 cm$^{-1}$ could however not fit with this explanation. However, it could be assigned to combination modes of the TA, ZO, LA, TO and LO phonon branches. Then, if we follow the 20° twist angle hypothesis, bands at 2868, 2974 and 3057 cm$^{-1}$ could be attributed to 2TO, LO+TO and 2TO overtones. In this respect, less intense bands at 1788, 1930, 2196, 2206 and 2275 cm$^{-1}$ could be attributed to LO+ZO', 2LA, TO+ZO, LO+TA, and LO+ZO, respectively. ZO' related band is predicted to fall in strong Si combination bands, explaining why it was not detected. The 20° twist angle explanation could be a quite good candidate, but 10 remaining bands could not be understood as overtones or combination of the previous modes related to the 20° twist interface. Moreover, bands at 1449 and 1528 cm$^{-1}$ were systematically detected, whatever the amount and the nature of FLG/MLG studied. Therefore, this explanation must be ruled out, except if an unknown mechanism triggers a systematic 20° twist angle somewhere in the material under the laser beam. Some other hypotheses are possible: a fourth one involves pure electronic resonance Raman process due to stacking effect on the electronic structure. Calculations made for few layers, organized in the rhombohedral order, display bands in ranges close to ≈1500, ≈2500, ≈3200 cm$^{-1}$ [89]. Twist interface of one layer with N=1, 2 or 3 other layers have also been investigated by the same group, showing other rich spectroscopic structures [60]. However, all these signatures are much broader than the ones observed in our case, which excludes this hypothesis. A fifth hypothesis is based on a pure phonon cause, involving turbostratic disorder. Thus, in [90], the authors focus on rotationally faulted CVD MLG, with up to 17 bands that have been found more or less equitably distributed between 1100 and 3200 cm$^{-1}$, due to double resonance mechanism, like the D, G and 2D bands. That hypothesis could be a satisfactory one, even if the authors count the TS bands in that 17 bands and that they do not fall at same location and are much broader than our bands. Still related to rotationally faulted stacked layers, another study, focused on flakes MLG with up to N=4 layers this time [56], reports on new bands rising in between the D and the G bands, typically close to 1470 cm$^{-1}$. Each band is related to a value of a twist angle for a given interface for 4 layers like 3+1, 2+2, 1+1+1, or 2+1+1. A sixth hypothesis involves SW and DV defects, reported to influence the vibrational spectrum of graphene [91]. SW alone lead to Raman bands in the range 1000-1200 cm$^{-1}$, close to 1600 cm$^{-1}$ (that could overlap with the D' band) and close to 1820 cm$^{-1}$. DV alone lead to Raman bands close to 560 cm$^{-1}$, in the range 1100-1300 cm$^{-1}$, at 1440 cm$^{-1}$, in the range 1516-1557 cm$^{-1}$, and in the range 1606-1702 cm$^{-1}$. Most of the bands observed below 1800 cm$^{-1}$ are close to the frequencies given by these calculations. None of hypotheses listed above is completely satisfactory, but the most promising one is related to the edge phonons. Note that these 23 bands are very intense not only for *od* - FLG but also for cr – FLG. Moreover, they are more intense for cr - FLG than for pt-FLG and *od* - FLG (see Fig.1-j - l). Thus, we could conclude that these bands are related to the degree of crumpling of the sample: the more it is crumpled, the more edge states are present under the Raman spot, leading to higher intensities. Then, 633 nm Raman microscopy could be used as an efficient tool to control that transferred graphene is done properly. This wavelength sensitivity could be due to a resonance between the laser wavelength and a vHS separation between conduction and valence bands. However, the ≈2 eV corresponding to 633 nm should involve a twist angle of about 8°, referring to the literature on tBLG [71]. This is far from the 20° that could have been involved to explain some of the bands found in the A and B spectral regions in the frame of another hypothesis. It means that if the resonance with vHS is involved here, it needs to be understood, including local structure induced by the crumpling, far from two perfect flat twisted layers of graphene.

One can ask about the possibility to observe Kohn anomalies and how they play a role in the rise of the 23 bands. Kohn anomalies in phonon dispersion at the gamma and K points in graphene are due to the breakdown of the Born-Oppenheimer approximation. As far as we know, these anomalies (kinks in the dispersion) change G and 2D band position and width in function of doping: $\sigma_G$ vary in the range 1580 to 1584 with doping up to $10^{13}$ cm$^{-2}$ whereas G band width is diminished by 6 cm$^{-1}$. To the best of our knowledge we have not found any work reporting Kohn anomalies for other phonon branches at the gamma and K points. Just below, we wonder if we could observe Kohn anomalies for the 23 new bands. In figure S4 we display a zone of the cr-FLG sample recorded at 633 nm. We found that:

- G and 2D band can be shifted by something like 10 cm$^{-1}$ depending on the zone of the sample.
- When G shifts, 2D shifts as well but $E_1$ and $E_2$ does not.
- However, $E_1$ and $E_2$ band intensities are high at the edges of the zones where 2D and G shift.

These observations cannot allow us to prove the existence of Kohn anomalies related to the new bands ($E_1$ and $E_2$ being the more intense). However, this observation confirms these bands are related to edges. In figure S5-a we also controlled that the polarization behavior does not affect these bands, whereas the Raman intensity of the 2D band is maximum when the excitation and detection polarizations are parallel and minimum when they are orthogonal, whereas that of the G band is isotropic.

On one hand, we demonstrated that the 23 new bands observed in the Raman spectra of FLG and MLG at 633 nm could be explained by the edge phonons hypothesis. On the other hand, bands at 1450 cm$^{-1}$ and 1530 cm$^{-1}$ are also very often found in literature when depositing molecules with structures and molecular symmetry similar to that of graphene on graphene surface to obtain a Graphene Enhancement Resonant Spectroscopy (GERS) effect [92] [93]. We ~~have to~~ must mention that for most of these molecules, the corresponding Raman spectra are very similar to the one we present on figure 4: bands in the A, B, C and E domains are also observed in these systems, with some shifts compared to our data. As it is not possible that we have introduced accidentally the molecular species reported in literature (like phtalocyanine), the fact we both observe similar spectra with different causes involved is puzzling and has never been reported until now to the best of our knowledge. One possibility could be that the GERS signal observed in literature was not due to a molecule of interest, but to edges activated by the molecule adsorbed on the surface. This hypothesis may be ruled out as very similar spectra are observed on other 2D substrates (h-BN and $MoS_2$). Another more reasonable hypothesis could be that the molecules leading to the GERS effect organize like nanoribbons, may be because of the high power laser needed to record spectra, explaining why spectra of these molecules contain the edge related bands which are very close from one kind of molecule to another [94], and then to edge phonons.

*4.4 Heterogeneity of wrinkles/crumples create 2D defects influencing the Raman spectrum*

Figure 5 deals with intensities and intensity ratios of some band of interest for *od*-FLG, at 633 nm, where the 23 new bands are seen. Fig. 5-a displays the silicon band intensity variation that is related to the geometry of the surface so that we could recognize the shape of the disks mentioned earlier. Figure 5-b displays the logarithm of the G band height. As observed previously in figure 1 I and h, graphene had been ripped during the transfer and remained on the disk edges only. G band intensity varies by one to two orders of magnitude depending on the location (closer to 2 orders of magnitude for *od* - FLG and one order of magnitude for *nd* - FLG). TEM made on a thin foil obtained by direct FIB cut of *od* – FLG zone revealed that the intensity increase is not caused by an accumulation of graphene layers during the sample transfer (see SI). Then, intensity enhancement is due to a modification of an electronic resonance, involved by crumpling, as mentioned previously. This should be investigated in more detail in the future, both experimentally and theoretically. In figure 5-c and d we display relative height ratios of respectively the D and 2D bands compared to the G band. $H_D/H_G$ vary from ≈ 0.02 to 1 whereas $H_{2D}/H_G$ vary from ≈ 0.5 to 5. Both are found to vary by one to two orders of magnitude at a lateral scale lower than the micron. For $H_D/H_G$, the traditional interpretation is that it is related to the number of defects whereas for $H_{2D}/H_G$ it is related to the quality of the graphene layers, a value of 2-3 being thought of good quality for $H_{2D}/H_G$ (see references in the introduction). Here, when $H_D/H_G$ is found high (i.e. "high amount of defects"), $H_{2D}/H_G$ at the same location on the sample is also find high (i.e. "good quality"). This is counter intuitive, as one expects goodquality for the low amount of defects. Interpreting relative intensity ratio ~~has~~ must to be done carefully as a given process could affect only one of the bands leading arbitrarily to a high variation of the relative intensity ratio, resulting in a wrong conclusion. Then, we should consider the absolute height of the bands. However, as the surfaces are not flat, local substrate topography should affect absolute intensity. The best way to correct substrate topography effects on the intensities is to consider relative intensity ratio involving the substrate signal. In figure 5-e to g, we display $H_G$, $H_D$, and $H_{2D}$ relatively to the silicon band heights. Sub micrometric heterogeneities are seen, and at the location where $H_D/H_G$ and $H_{2D}/H_G$ are low, it appears this is because $H_G$ is enhanced. A white arrow points the location where the G band is maximal and does the same for other bands to show it is spatially shifted. These maps prove that a mechanism increasing the G band intensity is at play here. On the other hand, both the D and 2D bands are found more intense in the same locations, different from the location where the G band is found more intense. Then, the mechanisms leading to the D and 2D bands intensity enhancement seem to be comparable, but different from the G band one. An order of magnitude increase in G band intensity could be reached in the framework of quantum interference effects due to charge doping of the order of few $10^{13}$ cm$^{-2}$ [95]. However, our charge doping could not be higher than few $10^{12}$ cm$^{-2}$, as deduced from Fig.3-a. Then, the doping could not be at the origin of this enhancement. Again, we have another indirect proof that local crumpling of FLG/MLG is responsible for such enhancement and needs to be investigated in more detail from first principles. We guess that this enhancement may have the same kind of origin as the enhancement found for tBLG and FLG submitted to environmental effects [32]. Figure 5-h displays the height of the band at 1448 cm$^{-1}$, the most intense of the 23 ones. $H_{1448}/H_{Si}$ behaves as the D and 2D bands, with maximum values found at the same locations, and not as G band. Then, in the resonance process leading to the 1448 cm$^{-1}$ band (and the 22 other bands), as the intensity of that band increases where the intensity of the D and 2D bands increase, and as these bands involve Dirac cones with a double resonance mechanism, a double resonance mechanism should also be at play. In other words, using spatial separation (i.e. Raman maps) suggests that the 23 new bands may have some common ingredients as the bands supposed to be formed by the double resonance mechanism. As we only observe these bands at 633 nm, contrary to the double resonance bands, it implies that not all the physical ingredients at play are the same.

Figure 6 displays a useful plot allowing to disentangle point and linear defects in SLG(referred here as 0D and 1D defects, respectively) by means of Raman microscopy[41]. Similar plot has been used in the past independently for both SLG and disordered carbons [37]. Here we show that this plot could also be useful to detect '*intensity anomalies*', which could help characterizing something else than '*just*' the amount of defects. This plot is not easy to understand, and our work contributes to its interpretation by visualizing how vHS present in wrinkles/crumples could affect the intensities. $A_D/A_G$ area ratio, corrected by the $E_L^4$ factor allowing not to consider the laser energy influence, are plotted in function of the G band width. Canonically, $A_D/A_G$ and $\Gamma_G$ are both known to be sensitive to the amount of defects. Moreover, $\Gamma_G$ is known to be sensitive to electron-phonon coupling, and then doping could influence it. The purple zone is where samples with a mixture of 0D and 1D defects are found in SLG. We did not put corresponding experimental data for clarity. Turbostractic graphite from literature, as well as our disordered C samples mimicking turbostratic graphite, are spread in an orange region, with a strong overlap with the purple region. Conversely, flattened carbon nanotubes reported to exhibit a strong D band enhancement without 0D defect introduction [96], display a vertical shift in the plot. Here we compare our FLG and MLG with all these reference samples to discuss different effects. Transferred FLG are found close to the purple region. *fd*-MLG are found in the purple zone for one half of the samples, with similar behavior than some turbostratic graphite, and outside for the other 50%, with a $A_D/A_G$ value diminished by a factor of ≈ 2. *od* – FLG measurements obtained at both 514 and 633 nm are found in the continuity of the cloud of points of *fd*-MLG. 514 and 633 nm data are at the same location in the plot. For comparison, we know that for tBLG, the two set of points should not be at the same location in the plot: the G band intensity changes resonantly depending on the twist angle –by two orders of magnitude- due to resonance with a corresponding vHS. Then, the interpretation for od-FLG strengthens our guess at this step: if there is no wavelength dependency whereas we use the wavelength corrected intensity, it means we are not looking at a unique resonance for a given twist angle under the laser beam. On the contrary, we probe a distribution of twist angles, each being resonant for one wavelength, (and then a twist angle with its corresponding vHS). The peculiar values of $A_D/A_G \times E_L^4$ are thought to be due to wrinkles/crumples that behave locally as tBLG, but with a distribution of twist angles under the laser beam due to heterogeneity in their texture. We propose to consider it as a '2D defect'.

Coming back to the general aspect of the plot in figure 6, both fd-MLG and od-FLG draw a new zone, the blue zone, with some overlap with the purple and the orange zones. This blue zone is very important because it allows to identify the '2D defects'. As a confirmation, *wr* - FLG is found to fall exactly in the middle of the blue zone, not so far from cr - FLG. To resume this graph: $A_D/A_G \times E_L^4 \approx 10$ seems to be a median value important to use as a reference. A value higher than 10 seems to indicate a D band enhancement caused by defects and/or structural distortion. A value lower than 10 indicates a G band enhancement caused by a high amount of edges under the laser beam. Most of our FLG/MLG have $\Gamma_G$ values falling in the 10-20 cm$^{-1}$ range, far from the one present in turbostratic graphite samples. This 10-20 cm$^{-1}$ range is the range explored for twisted BLG [72]. D band intensities are not reported in the literature when considering twisted BLG, but G band intensity variation with the laser used are reported. We present this amplitude as a grey vertical arrow in figure 6. This amplitude is caused by a resonance between the laser energy and the electronic structure modified by a vHS resulting from a given twist angle. One can see that the G band increase we present for our sample is compatible with the one existing for tBLG in the literature. Moreover, we think that the G band intensity we measure is mainly due to only one interface with only the good twist angle leading to the resonance with the corresponding vHS. More experiments controlling finely the number of layers, the texture of the wrinkles/crumples are then necessary to conclude.

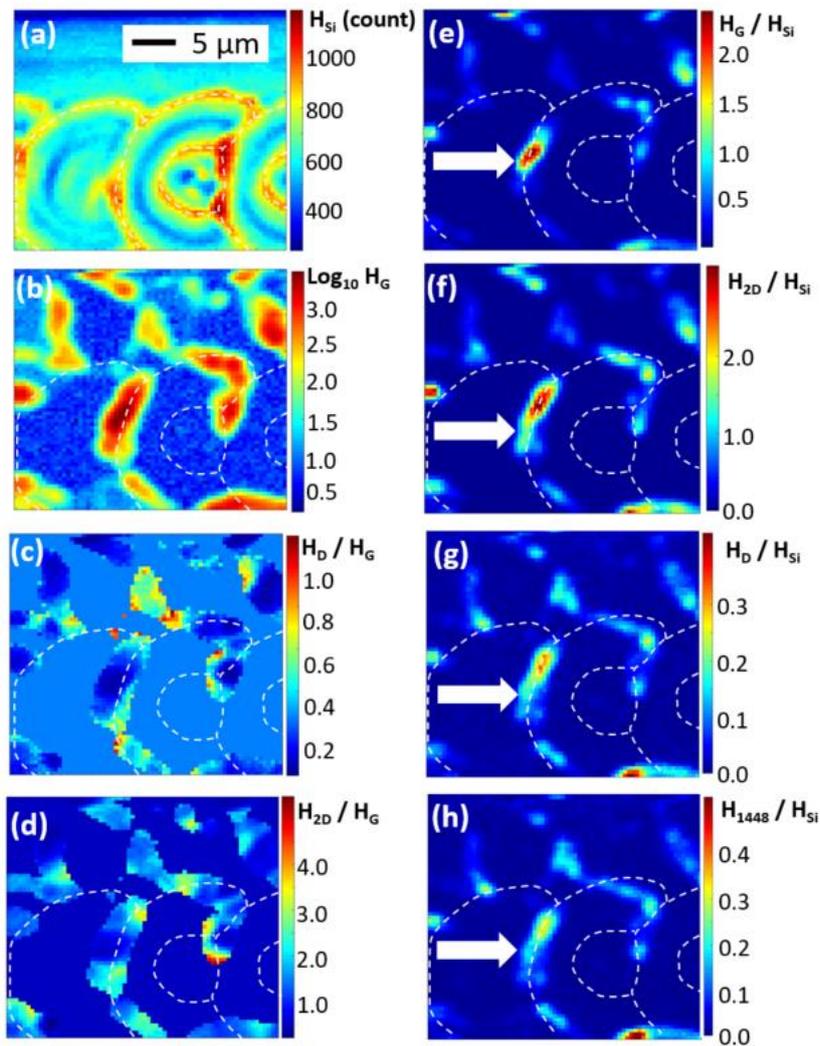

*Figure 5.* 633 nm-Raman intensity and intensity ratio maps of od-FLG. (a) $H_{Si}$. (b) $Log_{10}$ ($H_G$). (c) $H_D/H_G$. (d) $H_{2D}/H_G$. (e) $H_G/H_{Si}$. (f) $H_{2D}/H_{Si}$. (g) $H_D/H_{Si}$. (h) $H_{1448}/H_{Si}$. $H_{Si}$ corresponds to the maximum height of the second order Si band close to 950 cm$^{-1}$. The mode à 1448 cm-1 correspond to the A2 band presented previously. White arrow in the right column indicates the same position in each subplot (e-h).

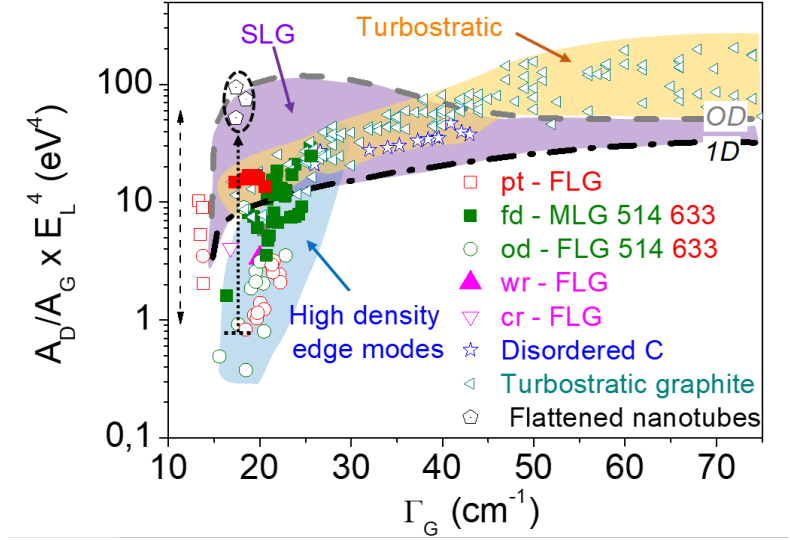

***Figure 6.*** *Intensity anomaly detection and defect identification plot using $A_D/A_G$ versus $\Gamma_G$. Turbostratic graphite data are from [79]. Flattened nanotubes data are from [96]. Note that the horizontal dot line and the vertical dot arrow related to flattened nanotubes highlight the enhancement, the horizontal line being the limit of detection of the D band when nanotubes are not flattened. The vertical grey arrow display the amplitude of G band intensity variation for twisted bi layer graphene [71].*

## 5. Conclusions

By studying with Raman microscopy a variety of transferred few and multi layer graphene, we reinvestigated its interpretation and usefulness. We introduced crumples and wrinkles on CVD graphene by 3 ways: by adapting the wet transfer process, by modifying the starting material with an AFM tip to cut and fold, and by using a corrugated laser machined Si wafer substrate. Using in correlation electron microscopies, we have shown that this crumpling/wrinkling introduces a high density of edge modes under the laser beam, which leads to the emergence of 23 resonant bands, only observed at 633 nm. By using a cut and fold method on three starting materials, with different strain and doping levels, we have shown that the folded multilayers obtained keep the memory of the initial material. We use this result to show that an adaptation of the $\omega_{2D}$ Vs $\omega_G$ plot is useful for the study of twisted bi layer graphene. We also have shown that we could use the $A_D/A_G \times E_L^4$ Vs $\Gamma_G$ diagram to detect the influence of the crumpling and wrinkling that modifies the standard double resonant bands' intensity, introducing the notion of '2D defects'. Both this diagram and the rise of the 23 bands could be used to optically qualify the multi layer graphene surface quality more efficiently. We discuss the possibility that the overall spectrum is dominated by the Raman diffusion of only one interface which display the good twist angle leading to the good resonance match with the resulting van Hove singularity.

**Acknowledgements**

A part of microfabrication processes was performed in PLANETE cleanroom facility (CINaM, Marseille).  We thank Dr. Igor Ozerov for technical support. Part of this research has been granted by Region Sud (France). This work received support from the French government under the France 2030 investment plan, as part of the Initiative d'Excellence d'Aix-Marseille Université - A*MIDEX." (*AMX-20-IET-015).* CP wants to acknowledge French ANR for future helpful fundings. We thank Pascale Roubin for fruitful discussions.

**Supplementary material**

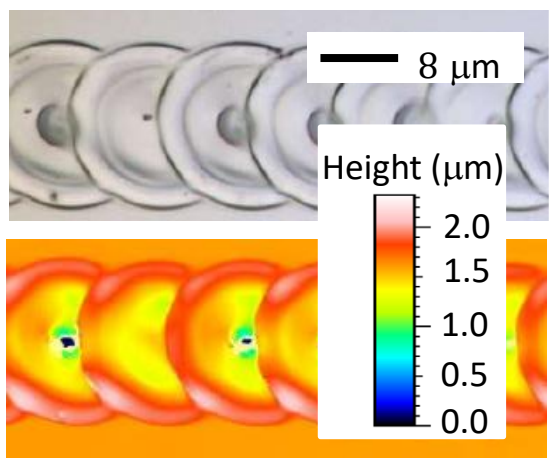

**Figure S1.** Confocal topography of silicon disks used

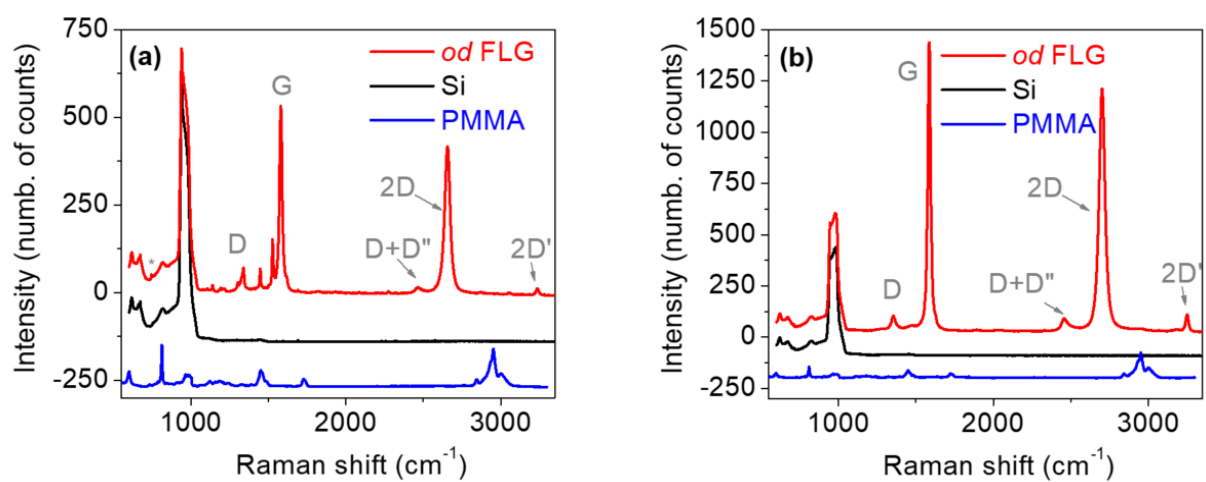

**Figure S2.** Raman spectra of Si, od FLG and PMMA at 633 nm (a) and 514 nm (b).

| Labelling in this study | Wavenumber (cm$^{-1}$) | Harmonic combination |
|---|---|---|
| D | 1328 | |
| G | 1580 | |
| D' | 1618 | |
| TS1 | 1825 | |
| TS2 | 1941 | |
| D+D" | 2463 | |
| 2D | 2655 | |
| 2D' | 3236 | |
| $A_3$ | 678 | |
| $A_2$ | 746 | |
| $A_1$ | 951 | |
| $B_3$ | 1105 | |
| $B_1$ | 1139 | |
| $B_5$ | 1181 | |
| $B_2$ | 1193 | |
| $B_4$ | 1212 | |
| $C_2$ | 1302 | |
| $C_1$ | 1339 | |
| $E_2$ | 1448 | |
| $E_1$ | 1528 | |
| $A_3+B_3/A_3+B_1$ | 1790 | 1783/1817 |
| $A_2+B_1$ | 1885 | 1885 |
| $2A_1$ | 1932 | 1902 |
| $A_1+B_3$ | 2016 | 2056 |
| $A_1+B_1$ | 2089 | 2090 |
| $E_2+A_3$ / $A_1+B_2$ | 2125 | 2126/2144 |
| $E_2+A_2/E_1+A_3$ | 2204 | 2194/2206 |
| $E_1+A_2/2B_1$ | 2274 | 2274/2278 |
| $2E_2$ | 2867 | 2896 |
| $E_1+E_2$ | 2973 | 2976 |
| $2E_1$ | 3054 | 3056 |

***Table S1.*** *Raman band bestiary in on disk MLG with 633 nm.*

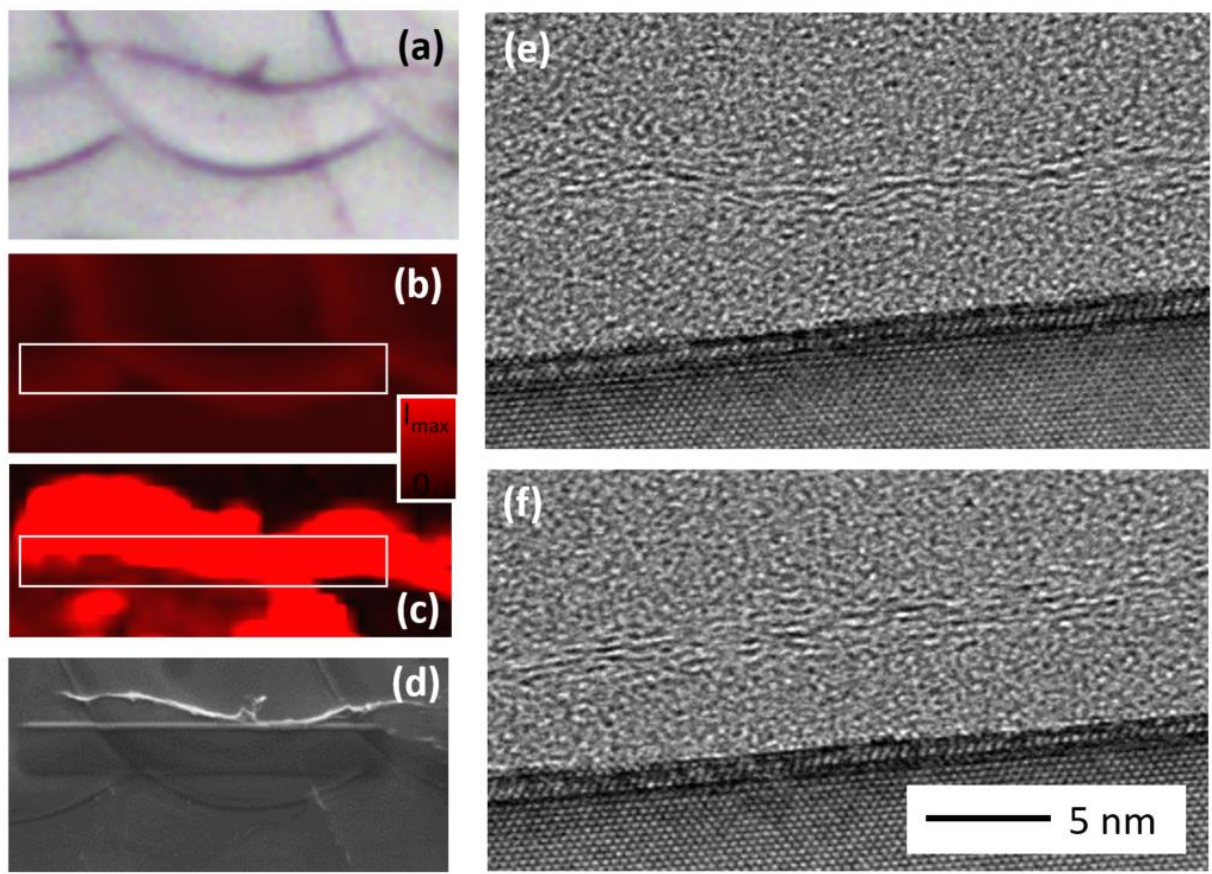

*Figure S3.* HRTEM measurements of on disk-MLG. (a) Optical microscope view. (b) Second order silicon band intensity map. (c) 1448 cm$^{-1}$ band intensity map. (d) SEM view, with the protective layer deposited before cutting the foil by FIB. (e-f) Typical HRTEM views. Scale bar for b and c is the same.

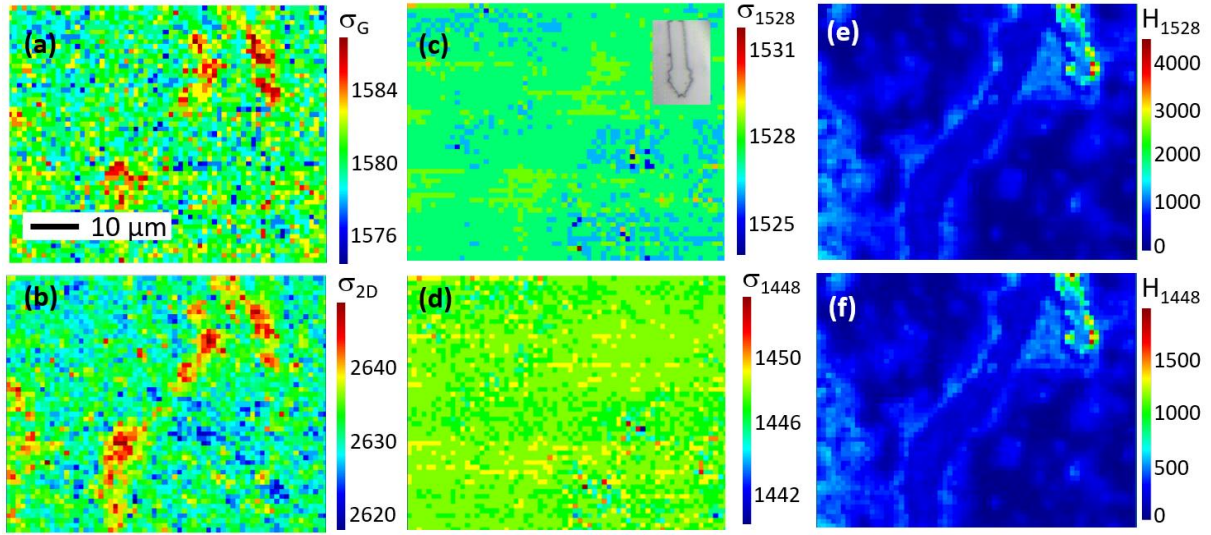

***Figure S4.*** *633 nm Raman maps of cr-FLG. (a, b) G and 2D band positions. (c, d) $E_1$ and $E_2$ band positions. (e, f) $E_1$ and $E_2$ band height. Inset in figure c is an optical image at the same scale showing an arrow patterned by lithography. No other optical counterpart could be seen in the field.*

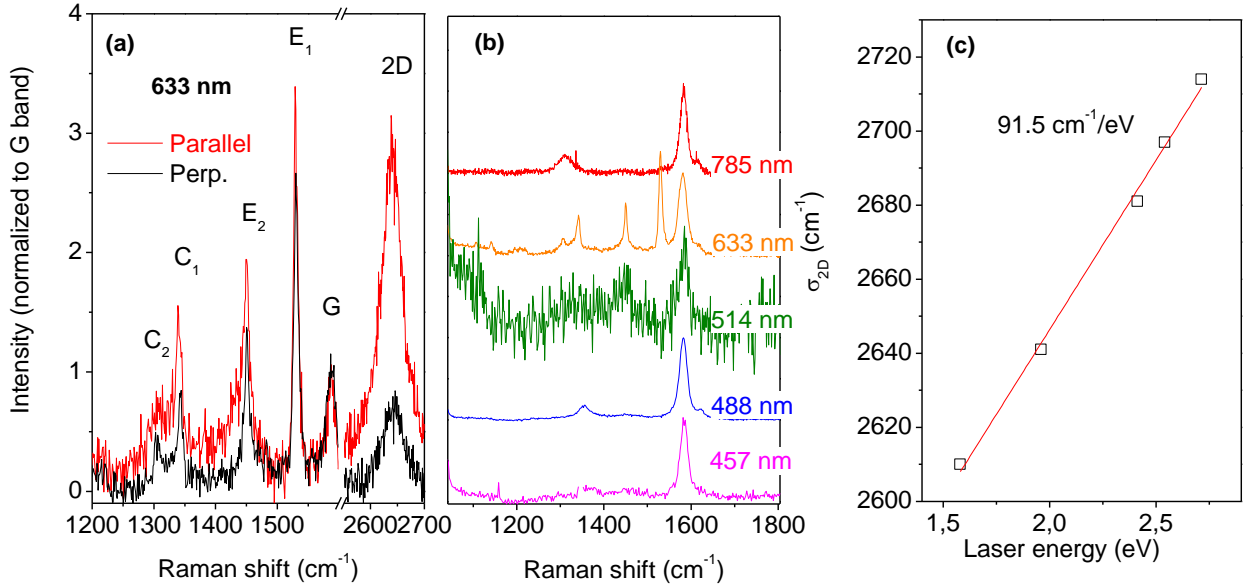

***Figure S5.*** *Cr-FLG. (a) Polarization study: excitation and detection polarizations parallel and perpendicular. (b) Multiwalength analysis. (c) 2D band dispersion.*

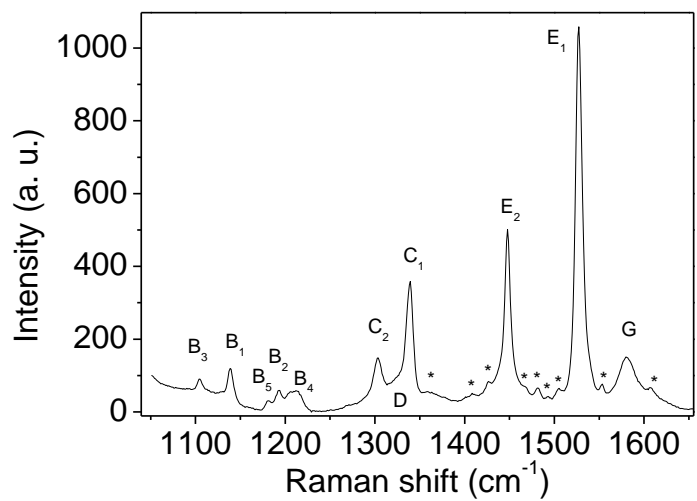

***Figure S6.*** *Raman spectra of cr-FLG summing the most intense parts of figure S4. Recorded at 633 nm. Stars are very weak bands that are seen only when the A, B, C and E bands are very intense and integration time overestimated.*